# How Segmental Dynamics and Mesh Confinement Determine the Selective Diffusivity of Molecules in Crosslinked Dense Polymer Networks

Baicheng Mei [a,d], Tsai-Wei Lin [c,d], Grant S. Sheridan [a,d], Christopher M. Evans [*a,c,d], Charles E. Sing [*a,c,d] and Kenneth S. Schweizer [*a,b,c,d]

[a.] *Department of Materials Science, University of Illinois, Urbana, IL 61801, USA*
[b.] *Department of Chemistry, University of Illinois, Urbana, IL 61801, USA*
[c.] *Department of Chemical & Biomolecular Engineering, University of Illinois, 61801, USA*
[d.] *Materials Research Laboratory, University of Illinois, 61801, USA*

[*] [kschweiz@illinois.edu](kschweiz@illinois.edu)   [*] [cesing@illinois.edu](cesing@illinois.edu)   [*] [cme365@illinois.edu](cme365@illinois.edu)

# Abstract

The diffusion of molecules ("penetrants") of variable size, shape, and chemistry through dense crosslinked polymer networks is a fundamental scientific problem that is broadly relevant in materials, polymer, physical and biological chemistry. Relevant applications include molecular separations in membranes, barrier materials for coatings, drug delivery, and nanofiltration. A major open question is the relationship between molecular transport, thermodynamic state, and chemical structure of the penetrant and polymeric media. Here we address this question by combining experiment, simulation, and theory to unravel the competing effects of penetrant chemistry on its transport in rubbery and supercooled polymer permanent networks over a wide range of crosslink densities, size ratios, and temperatures. The crucial importance of the coupling of local penetrant hopping to the polymer structural relaxation process, and the secondary importance of geometric mesh confinement effects, are established. Network crosslinks induce a large slowing down of nm-scale polymer relaxation which greatly retards the rate of penetrant activated relaxation. The demonstrated good agreement between experiment, simulation, and theory provides strong support for the size ratio variable (effective penetrant diameter to the Kuhn length which quantifies the stiffness of the polymer backbone) as a key variable, and the usefulness of coarse-grained simulation and theoretical models that average over Angstrom scale chemical details. The developed microscopic theory provides a fundamental understanding of the physical processes underlying the behaviors observed in experiment and simulation. Specifically, both local steric caging of penetrants by polymer segments in the first solvation shell, and the causally coupled effect of the longer-range collective elasticity of the crosslinked network, resist penetrant hopping in distinctive manners. Penetrant transport is theoretically predicted to become even more size sensitive in a more deeply supercooled regime not probed in our present experiments or simulations, which suggests new strategies for enhancing selective polymer membrane design.

**Introduction**

The permeation of atoms, molecules and nanoparticles of variable size, shape, and chemistry (referred to here as "penetrants") through a dense polymeric media (liquid, glass, crosslinked permanent or dynamic rubber network, thermoset) is a rich fundamental scientific problem that is broadly relevant in materials, polymer, physical and biological chemistry [1-14]. Understanding how the many physiochemical factors and thermodynamic state control penetrant activated mass transport is critical in many applications including gas, water, and organic molecule separations in membranes [2, 8, 9, 11, 15-18], self-healing based on microcapsules [19-21], ion and solvent transport in biological and polymeric materials [9, 22, 23], barrier materials for coatings [20, 21, 24], drug delivery [5, 25], and nanofiltration [26, 27]. In particular, it underlies a massive chemical separations industry traditionally based on costly and often inefficient distillation methods estimated to constitute 10-15% of the world's energy consumption [17]. This has motivated an intensive search for alternative membranes that filter molecules based on their size, shape and/or interactions with the matrix. A common paradigm is to control the size of membrane pores based on materials such as metal organic frameworks, covalent organic frameworks, and zeolites.[28-31] Yet, these materials are typically mechanically brittle and difficult to process and deploy at scale. Glassy amorphous polymers are more viable for gas separations [2, 32, 33], but less size-selective, and ineffective for separations of larger molecules such as polyaromatics which are excluded in such low "free volume" solid materials. Rubbery membranes exploit equilibrium solubility as a selectivity strategy [34], but transport is relatively fast and differences in solvation are generally not as sensitive to molecular structure as in materials designed to exploit the activated nature of penetrant diffusion (e.g., polymer glasses). Overall, a major conceptual bottleneck remains due to our still-nascent understanding of the relationship between molecular transport, thermodynamic state, and the chemical structure of the separating species and the polymeric membrane.

Here, we consider non-gaseous molecular penetrant diffusion in solvent-free, crosslinked polymer networks. Even in the dilute limit, understanding how molecular aspects of the polymer control the absolute and relative rates of penetrant diffusivity remains open since activated processes in condensed media can be exceptionally sensitive to a multitude of local physical and chemical factors that characterize the penetrant and polymer species. Recent theoretical work for polymer melts has predicted [35] that in the supercooled regime and approaching the glass transition temperature ($T_g$), the sensitivity of penetrant hopping to polymer structural relaxation is strongly amplified. Exploitation of this behavior can be an effective means for developing new selective membranes, which we hypothesize can be even further improved if polymer melts are covalently crosslinked.

Our main goal is to study and understand molecular penetrant diffusion in the dilute limit in supercooled crosslinked polymer networks down to $T_g$. Chemical crosslinks induce additional dynamical constraints on penetrant diffusivity via two *qualitatively* distinct processes: (i) slowing down of nm-scale polymer segmental relaxation in a highly temperature-dependent manner, and (ii) entropic (virtually temperature-independent) geometric mesh obstruction of penetrant diffusion, effects which are both enhanced with increasing crosslink density. Here we clarify, for the first time, the competition and inter-relationship between these two effects on penetrant transport over a wide range of temperatures, network mesh sizes, and penetrant sizes. This is a major challenge that classic phenomenological models [9, 22, 36-39] cannot address. Moreover, most prior discussions of mesh obstruction effects are based on rigid frameworks or theoretical models for larger nanoparticles in a high temperature rubber [40-42].

Our results are of broad relevance towards addressing the physicochemical principles underlying points (i) and (ii) above and are based on a coordinated experimental, simulation, and theoretical effort. By experimentally selecting penetrant molecules of variable size and shape within a class of aromatics with similar dispersive (weak) attractions with the polymer matrix, we address how chemical crosslinks and activated segmental relaxation simultaneously conspire to determine the penetrant mass transport rate. The simulation and theory-based modeling purposefully ignores chemically specific penetrant-polymer attractions and adopts coarse-grained spherical penetrant and polymer network models that average over Angstrom scale details. This simplification is motivated not only by practical computational and theoretical considerations, but our desire to understand if such fine chemical details are strongly self-averaged in the determination of long-time penetrant diffusion constants. Encouraging support for this concept has been recently established based on a combined theory-experiment analysis [43] of a large set of atomic and molecular penetrant diffusion data in chemically diverse molecular and polymer liquids over a broad range of temperatures down to the glass transition. Confrontation of the new experimental data in crosslinked networks presented here with our theoretical and simulation results shows favorable agreement. This allows us to draw multiple important conclusions that illustrate the generality of the competing physical mechanisms, which will be of major value for the experimental design of selective membranes and the usefulness of simulation and theory for the penetrant transport problem.

**Materials and Methods**

All technical details of the experiment, simulation, and theory can be found in the Supporting Information (SI) and our prior papers [44]. Here we sketch the essential elements.

Using fluorescence methods we experimentally measured at 23 ºC the diffusion constant of four penetrant dye molecules [N,N'-Bis(2,5-di-tert-butylphenyl)-3,4,9,10-perylenedicarboximide (BTBP), tert-butylated rubrene (TBRb), rubrene (RUB), and hydroxy functionalized nitrobenzoxadiazole (NBD-OH) of differing shape and size (**Figure 1a**) under dilute conditions in previously well-characterized [14] dense poly(n-butyl-acrylate) (PnBA) polymer networks (schematic of synthesis in **Figure 1b**; crosslink fraction $f_{cross}$). The penetrant van der Waals volumes were calculated using an atomic group contribution method [45, 46] (see Table 1). To isolate the effect of molecular shape, BTBP (rod-like with aspect ratio ~3.5) and TBRb (more rounded and plate-like) were selected since they have nearly identical space filling volumes. To consider the penetrant volume effect at roughly similar shape, we studied TBRb and RUB. The NBD-OH molecule was examined since it is more spherical than the other penetrants studied, has a significantly smaller volume, and contains a positive and negative charge distribution on the nitro group that is absent in the nonpolar TBRb, RUB and BTBP penetrants. Characterization data are given in the SI (**Figures S1-S3**).

Standard Molecular Dynamics (MD) simulations were performed using our previously validated coarse-grained model for PnBA networks (semiflexible chains [47-52] where each bead represents a single nBA monomer of size $\sigma^*$) with explicit crosslinks as the polymer matrix in which dilute spherical penetrants (diameter $d$) are dissolved (**Figure 1b**) [37, 39, 41, 42, 53]. The network mesh size is computed as the average distance between crosslink points, and the penetrant diffusion constant extracted from the long-time limit of its mean square displacement; see, e.g., **Figure S4**, and the SI for other technical details of the simulation [54-57].

Here we generalize to crosslinked networks the self-consistent cooperative hopping (SCCH) statistical mechanical theory of the activated spherical penetrant (diameter $d$) relaxation time previously developed for polymer melts [35]. The approach self-consistently predicts the variable degree that smaller scale polymer segmental motions are coupled with the penetrant activated hopping event. **Figure 1c** presents a schematic of the physical ideas which are based on coupled dynamic free energies for the penetrant and polymer matrix that define the forces on a moving penetrant and Kuhn segment (length $l_K$), respectively, a priori constructed from knowledge of the microscopic structure. The penetrant activated barrier hopping time and the Kuhn segment correlated motion are self-consistently determined [35]. **Figure 1c** also indicates how the polymers are coarse-grained to a tangent semiflexible chain of hard-core beads (diameter $\sigma$) with parameters that model PnBA, with network crosslinks (fraction $f_n$) modeled as regular pinning of immobilized beads along a chain [44]. All intermolecular interactions are hard-core repulsions, and the structural correlations required to quantify dynamical constraints are computed using the polymer reference interaction site model integral equation theory [35, 44, 58]. The hard-core model is related to thermal networks at 1 atm by a mapping procedure [44, 59, 60] using PnBA experimental equation-of-state data [43, 44]. The mean network mesh diameter is calculated as $a_x = \sigma/A f_{cross}^{1/2}$ where $A$ is calibrated for the PnBA networks and a linear connection between $f_n$ and $f_{cross}$ applies (see SI and Ref. [61]). The penetrant alpha process involves hopping over a local cage barrier ($F_{B,p}$) which is coupled with longer-range collective displacements of all Kuhn segments outside the cage characterized by an elastic barrier ($F_{el,p}$). The total activation barrier is the sum of these two barriers reflecting the coupled local-nonlocal nature of the penetrant relaxation process. The theory predicts the penetrant alpha time as a function of crosslink density, penetrant-to-matrix size ratio ($d/\sigma$), and temperature. SCCH theory does not explicitly account for mesh confinement effects on penetrant diffusivity.

Prior quantitative comparisons of the segmental alpha time results for the PnBA networks from theory, simulation, and experiment show excellent agreement.[44] Quantitatively different coarse-graining methods are used corresponding to bare bead sizes of $\sigma^* = 0.573$ nm in simulation [44] and $\sigma = 1.03$ nm in the theory, per Table 1. Both models adopt the experimental PnBA Kuhn length $l_K = 1.72$ nm [62]]. The temperature mapping in the theory quantitatively differs from how scaled temperature is converted to absolute temperature in simulations [44].

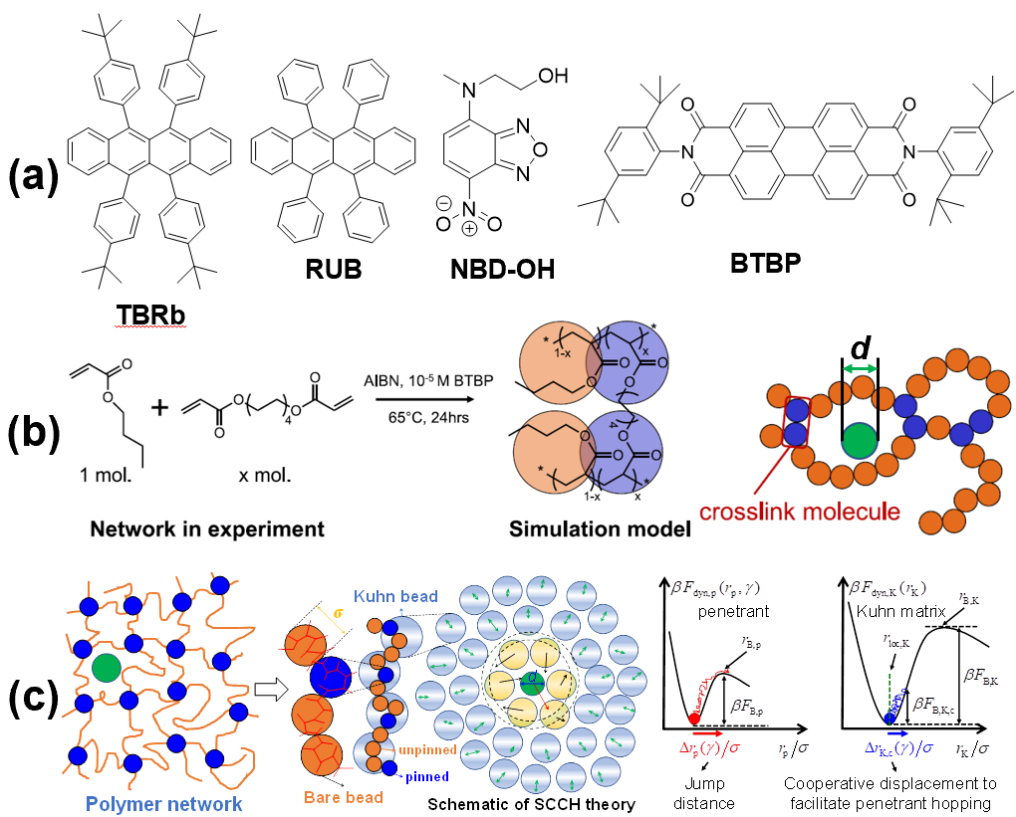

**Figure 1.** (a) Chemical structures of the penetrants studied. (b) Reaction mechanism for creating the experimental PnBA networks [14, 44] and the corresponding coarse-grained simulation model [44] where the crosslink (normal) segment is colored blue (orange). (c) Schematic (from left to right) of the polymer network model employed in the theory which based on the introduction of crosslinks via the regular pinning of beads along the chain [44], how polymers are coarse-grained to the Kuhn segment scale [35, 44, 58], and the physical ideas of SCCH theory [35, 63-65] based on coupled dynamic free energies for the penetrant and Kuhn segment displacements with a trajectory coupling parameter defined as $\gamma = \Delta r_p(\gamma)/\Delta r_{K,c}(\gamma)$ and relevant length and energy scales indicated. Note that although the normal and cross-linked beads are colored the same as those in (b), the beads or interaction sites of the semiflexible chain model is not identical to the meaning of a bead in the simulation model [44].

**Table 1.** Van der Waals volume $V$ of the four penetrants investigated [45, 46] and their corresponding diameter $d$ if treated as a sphere of equivalent volume [43]. Penetrant-to-matrix size ratio computed based on $\sigma = 1.03$ nm evaluated using $l_K = 2l_p - \sigma$, where the Kuhn length of PnBA melts $l_K = 1.72$ nm and the ratio of the persistence length to bare bead size in the theory is $l_p/\sigma = 4/3$.

|  | BTBP | TBRb | RUB | NBD-OH |
|---|---|---|---|---|
| $V$/Å$^3$ | 735.88 | 775.72 | 493.64 | 183.7 |
| $d$/Å | 11.2 | 11.4 | 9.81 | 7.05 |
| $d/\sigma$ | 1.09 | 1.11 | 0.95 | 0.69 |

**Conceptual Background and Overview of Core Results**

**A. Background**

Activated penetrant diffusivity in crosslinked networks is influenced by both polymer segmental dynamics and entropic geometric mesh confinement. In the qualitative, *but distinct* (see below), spirit of prior work [40-42], these two factors for a spherical penetrant of diameter $d$ enter its diffusion constant in a multiplicative manner as:

$$D_p \sim \frac{d^2}{\tau_{\alpha,p}} \, C^{-1} \exp(-bC^2) \equiv \frac{d^2}{\tau_{\alpha,p}} X, \tag{1}$$

Eq (1) relates penetrant diffusivity to a product of the rate of its *crosslink fraction (mesh size), size ratio, and temperature-dependent* elementary hopping time, $\tau_{\alpha,p}$ (the focus of SCCH theory), and a temperature-*independent* factor, $X$, associated with an entropic barrier for penetrant traversal through a polymer mesh. The latter depends on the degree of crosslinking quantified by a confinement parameter, $C = d/a_x$, where $b$ is a chemistry specific numerical factor originally [40, 41] suggested to equal unity corresponding to $X \equiv \exp(-C^2)/C$. Prior theoretical analysis [40, 41] addressed nanoparticle diffusion in rubbery crosslinked networks at a fixed high temperature in a regime with $C > 1$, very different from the present work. For such systems it is *not* the penetrant activated alpha time that enters as in Eq(1), but rather the *penetrant-independent* Rouse relaxation time of a polymer network strand which depends differently on temperature than the penetrant relaxation time predicted by SCCH theory. The expression $CD_p \propto \exp(-bC^2)$ does capture well simulations [41] for dilute spherical nanoparticles in *semi-dilute* high temperature rubbery polymer networks with a value of $b$ greater than unity (indirectly suggesting the importance of $\tau_{\alpha,p}$), despite the fact that $C$ greater than unity is not strictly obeyed for some of the simulated systems. In contrast, the experimental, simulation and theory work reported here for molecular penetrants in dry supercooled polymer networks indicate a different conclusion: crosslinking fraction dependent segmental relaxation is the most important process for activated penetrant diffusion. More generally, from a physical perspective using $\tau_{\alpha,p}$ in eq (1) seems appropriate since mass transport is initiated by a penetrant local hopping event.

For molecular transport the confinement parameter $C$ often does not exceed unity. This motivates us to briefly consider in the SI an alternative model of the effect of network meshes perhaps germane to such a weak ($C$ less than unity) confinement regime [42]:

$$D_p \sim \frac{d^2}{\tau_{\alpha,p}} \exp(-b'C) \qquad (2)$$

Here $b'$ is a system-specific constant, and we take $\tau_{\alpha,p}$ to be the same as in eq (1).

Our goal is to determine the relative and absolute importance of polymer segmental relaxation and confinement mesh obstruction on penetrant transport over a wide range of conditions. Unavoidably, directly separating the importance of the latter two factors from experimental or simulation measurements of $D_p$ is problematic. This is addressed using SCCH theory since it lacks the mesh effect, but it also can be included post facto per eqs (1) and (2). Before presenting our detailed results, we provide a high-level summary of our primary new scientific findings.

**B. Overview of Core New Findings**

The favorable confrontation of our experimental data with our theoretical and simulation results allows us to draw the following important conclusions. (1) The distinct physical effects of activated segmental relaxation and network mesh confinement affect activated penetrant diffusion to varying degrees, but our experimental and simulation results can be empirically *described* from either a locally activated penetrant hopping perspective where $T_g/T$ is the key variable and an "effective" Arrhenius behavior applies, *or* where the entropic mesh confinement parameter $C$ is the key variable. A theoretical understanding of this surprising "degeneracy of interpretation" is constructed. From a causality perspective, we find that the cooling and crosslinking induced slowing down of segmental relaxation is the dominant factor. (2) At fixed temperature, the penetrant diffusion constant decreases with its size as an inverse power law with a large exponent of non-hydrodynamic origin. (3) The good agreement of experiment with simulation and theory based on coarse-grained penetrant and polymer models indicates that local chemical detail even in tightly crosslinked networks is largely "self-averaged" in the determination of the long-time penetrant diffusion constant. (4) Our validated SCCH theory allows one to disentangle the relative importance the 3 different physical contributions to the activation barrier for molecular transport: (i) local caging by polymer segments in the first solvation shell of the penetrant, (ii) longer range small amplitude collective elastic displacements of polymer segments farther from a hopping penetrant, and (iii) network mesh confinement. Effects (i) and (ii) are intimately coupled in SCCH theory, but for small enough penetrants and/or hot enough polymer networks, the theory predicts the local caging effect (i) is dominant. (5) The theory makes new testable predictions that in a more deeply supercooled regime and/or for larger penetrants not probed in our present experiments or simulations, the rate of penetrant transport becomes even more size sensitive due to stronger coupling of its motion with polymer collective elasticity, a mechanism which holds promise for enhancing the selectivity of polymer membranes.

**Results**

Below we present our experimental, simulation, and theoretical results for penetrant diffusivity in the context of four core issues: (a) role of temperature and crosslinking fraction dependent polymer segmental relaxation, (b) role of penetrant-to-matrix size ratio, (c) role of entropic mesh confinement, and (d) mechanistic theoretical understanding.

**A. Role of Segmental Relaxation**

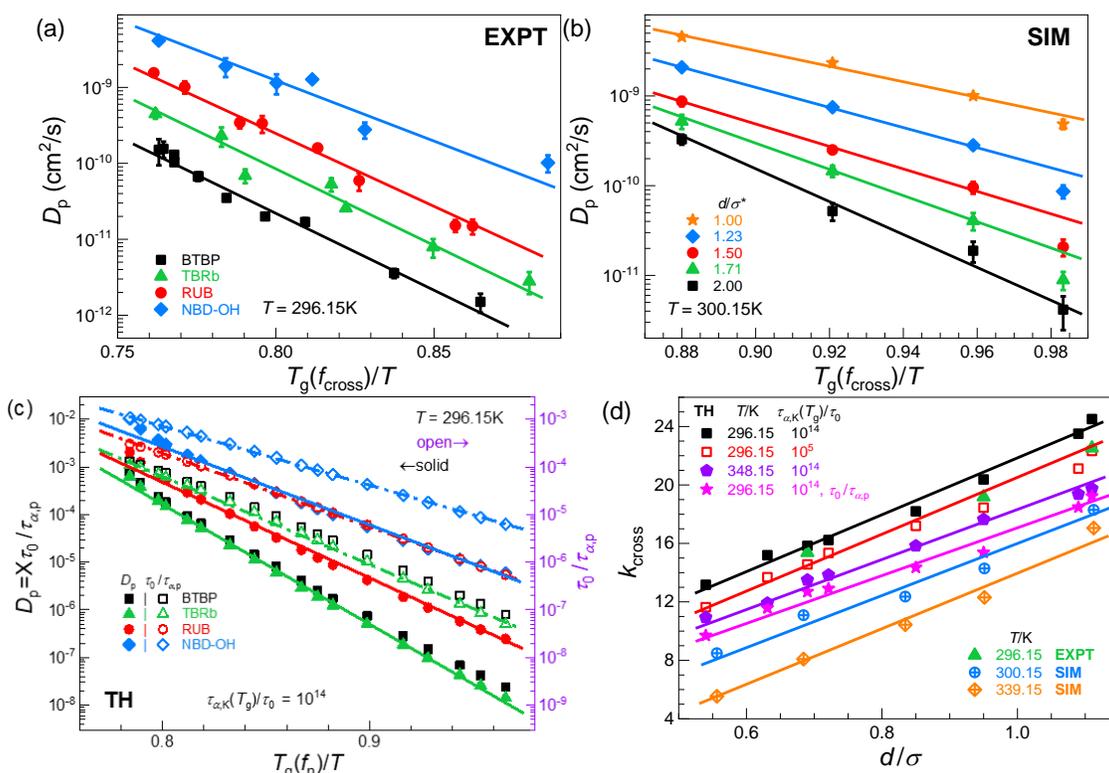

**Figure 2.** Relationship between penetrant diffusivity and the crosslinking fraction dependent network glass transition temperature at a fixed temperature. (a) Log-linear plot of the experimental (EXPT) diffusion constants as a function of $T_g(f_{\text{cross}})/T$ over a wide range of crosslink fractions at $T = 296.15K$ (23 $^0$C). (b) Same in (a) but from simulation (SIM) for 5 penetrants of variable diameters at $T = 300.15K$. Note the SIM diffusion constant data is vertically shifted down by 5 decades for comparison with the EXPT results. (c) Same as in (a) but for the theory (TH) with penetrant size ratios that mimic the experimental systems per Table 1; the theoretical $D_p$ (solid symbols with solid lines) is determined using Eq(1). Also shown is the predicted penetrant hopping rate, i.e., inverse alpha time, (open symbols with dash-dot lines, in units of $\tau_0$ corresponding to 1

ps per typical viscous liquids) with its y-axis indicated on the right-hand side colored in purple and over the same number of decades as that of the left y-axis. The $T_g$ criteria for EXPT, SIM, and TH in (a), (b) and (c) are the thermodynamic $T_g$ (see Table S1), $\tilde{\tau}_{\alpha,K}(T_g) = 10^5$ and $\tau_{\alpha,K}(T_g)/\tau_0 = 10^{14}$, respectively. (d) Effective Arrhenius activation energies deduced from EXPT, SIM and TH defined from the slope of the log $(1/D_p)$ versus $T_g(f_n)/T$ plot, as a function of size ratio $d/\sigma$ in a linear-log representation for different temperatures. The corresponding slopes for the log $(\tau_{\alpha,p}/\tau_0)$ versus $T_g(f_n)/T$ theoretical results are also shown. We note that in the simulation (theory) the bead size $\sigma^* = 0.573$ nm [44] ($\sigma = 1.03$ nm, see Table 1), and thus for comparison purposes the x-axis of the simulation results in panel (d) is scaled by a factor of 0.573/1.0.

Motivated by the hypothesis that the increase of the polymer segmental relaxation and $T_g$ with crosslink density is the dominant factor that determines the penetrant diffusion constant, **Figure 2a** plots the experimental data as a function of $T_g(f_{cross})/T$ at 23 °C (293.15 K). All penetrants show an exponential dependence on $T_g(f_{cross})$. The diffusion coefficient at fixed $T_g(f_{cross})$ decreases with increasing probe volume from NBD-OH to RUB to TBRb. Note that BTBP exhibits slower diffusion by a factor of 2-3 compared to TBRb despite having almost the same volume, although their apparent activation energies are nearly identical. This indicates a modest effect of molecular shape on mass transport and the relationship between diffusivity and $T_g(f_{cross})/T$. Overall, the dominant factor is the penetrant size, as seen in **Figure 2a** where the RUB and NBD-OH which have different slopes. **Figure 2d** presents the slopes of the activation plot of the data in **Figure 2a** as a function of the size ratio. The linear-log plot establishes an interesting power law behavior between penetrant diffusion constant and penetrant-to-matrix size ratio.

**Figures 2b** and **2c** show the corresponding simulation and theory results, respectively. Both capture the basic experimental trends, which are robust to increasing temperature beyond that studied experimentally (see **Figure S6**). This level of agreement seems remarkable given penetrants are modeled as spheres and the polymers are coarse-grained. The agreement also provides support for the idea that the penetrant to segment size ratio is the most important variable, as recently demonstrated for a wide range of chemically diverse penetrants in molecular and polymer melts [43]. Of course, use of a spherical penetrant model does not account for the modest differences between the diffusivity of BTBP and TBRb observed experimentally. We note that based on other theoretical models [40-42] that assume that entropic mesh confinement effects dominate the penetrant diffusivity with a crosslink fraction *independent* prefactor rate, the results in **Figure S6** would be the same as that of **Figures 2b** and **2c** since the confinement parameter is temperature independent. However, significant differences at different temperatures exist (especially the slopes per **Figure 2d** and absolute values), supporting the idea that activated segmental relaxation (via $\tau_{\alpha,p}$ in eqs. (1) and (2)) is of major importance for penetrant diffusivity.

A potentially important difference between simulations and the experiments and theory predictions is the disparity in time scales probed: the simulation $T_g$ is defined when $\tau_\alpha(T_g) \sim 100$ ns, dramatically shorter than in experiment and theory. Nevertheless, the overall good agreement between simulation, theory, and experiment can be understood from our prior work on neat polymer networks [44] where it was demonstrated that the vitrification criteria does not significantly change the *normalized T*-dependence of the network alpha time *in* the weak and moderately supercooled regimes. Moreover, the inset of **Figure S6** confirms that the *T*-dependence of the penetrant diffusivity similarly remains unchanged by the vitrification criteria adopted.

SCCH theory allows one to separate the effects of activated segmental relaxation and mesh confinement on penetrant mobility in networks. This is done in **Figure 2c** by plotting the theoretical $D_p$ with and without the factor of *X* in Eq (1). Both sets of theoretical results are consistent with an exponential dependence on $T_g/T$, with the *T*-dependence of $D_p$ computed including the factor *X* modestly stronger than computed if the mesh obstruction factor is ignored. This quantitative analysis suggests mesh effects are of minor importance for the penetrants and polymer networks studied relative to the consequences of crosslinking induced slowing down of polymer segmental relaxation.

A further consistency check between experiment, simulation, and theory is to examine the slope of log ($1/D_p$) versus $T_g/T$ in **Figure 2a**, **2b**, and **2c**, defined as $k_{cross}$. **Figure 2d** shows it increases with size ratio in an interesting logarithmic manner. Moreover, the experiments (green triangles) are consistent with theory and simulation for very different vitrification criteria. This is true even for the polar molecule NBD-OH that presumably interacts with the polymer in a more chemically specific manner. This further confirms that penetrant size is key in determining the activated diffusion of penetrant, and supports our hypothesis that Angstrom-scale chemical structural features are significantly self-averaged in the long-time diffusivity. At a finer quantitative level, simulation does underpredict the magnitude of $k_{cross}$, presumably due to the specific coarse-grained model employed and/or the fact it does not probe the more deeply supercooled regime. Theoretical predictions for $\tau_0/\tau_{\alpha,p}$ that neglect the mesh confinement contribution (no factor of $X$ in Eq (1)) lead to modestly smaller values of $k_{cross}$ that deviate more from experiments. Interestingly, one observes that the various results from the $k_{cross}$ versus $d/\sigma$ plot for $\tau_0/\tau_{\alpha,p}$ are parallel to their analogue for $1/D_p$. This indicates the mesh confinement contribution $X = \exp(-C^2)/C$ to $k_{cross}$ is effectively independent of $d/\sigma$ to leading order. This can be understood as a consequence of the barrier from entropic confinement, i.e., the slope of the $-\log(X) \sim C^2$ versus $T_g(f_n)/T$ plot, changes only weakly (by a factor of ~3.5-4.7) since we find $C^2 \propto T_g(f_n)$ (see **Figure S5** and relevant discussion in the SI).

Although changing temperature is not our primary focus, the magnitude of $k_{cross}$ at elevated temperatures has been examined using simulation and theory in **Figure 2d**. In both cases, heating leads to smaller values of $k_{cross}$ due to the reduced barrier for local penetrant hopping.

This finding relates to the recent work in polymer melts which established [43] that the dependencies of the penetrant diffusion constant on size ratio and temperature are nearly independent (barrier factorization), and the temperature dependence is determined by properties of the pure polymer matrix. Thus, at different temperatures the dependence of $k_{\text{cross}}$ on penetrant size is not expected change, as confirmed in **Figure 2d** by the parallel behavior of the data at different temperatures.

**B. Role of Penetrant to Polymer Segment Size Ratio**

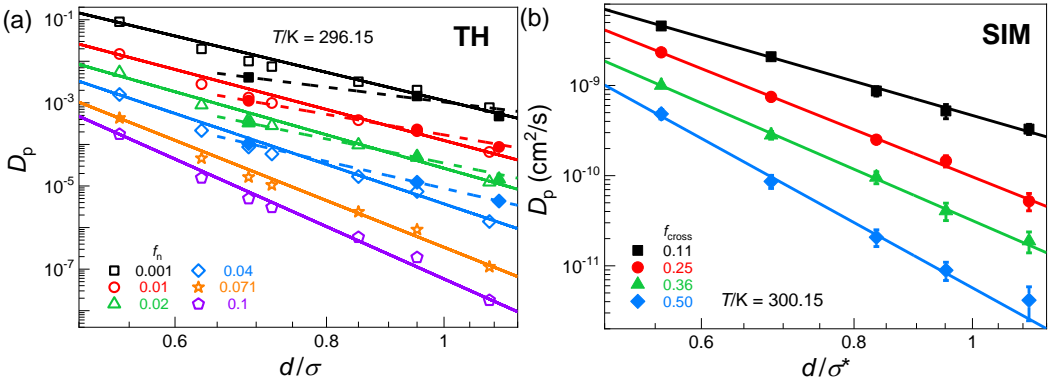

**Figure 3.** Penetrant diffusivity at various fixed crosslink densities as a function of penetrant-to-matrix size ratio $d/\sigma$ plotted in a log-log manner. EXPT data in (a) is the filled symbols vertically shifted up by a factor of 6 decades for comparison. TH results (a) are shown at $T = 296.15$K (open symbols), and (b) SIM results at $T = 300.15$K. A summary of the slopes from EXPT, SIM, and TH is given in **Figure S8.** As in **Figure 2d**, for the x-axes the SIM value of the size ratio variable is multiplied by the factor 0.573/1.03.

**Figure 3** presents theory and simulation results that establish how $D_p$ varies with size ratio over a wide range of crosslinking fractions at a fixed temperature. This aspect is relevant to the problem of isothermal selectivity of penetrant transport based on size. Given our findings above that $\log(1/D_p) \propto T_g(f_{\text{cross}})$ and the proportionality constant is a logarithmic function of size ratio $d/\sigma$, we conclude that (i) probe diffusivity decreases as a power law with $d/\sigma$, and (ii) activated segmental relaxation dominates the penetrant size dependence. Interestingly, the finding (i) is qualitatively the same as found in prior SCCH theory analysis for dilute spherical penetrants in semiflexible polymer chain melts [35, 43] which revealed a power law relation applies well over a wide range of $d/\sigma$ (including the large size ratio regime), and which remains robust even for systems containing weak attractions between the penetrant and matrix [35]. Here we find that when crosslinking is introduced the predicted power law relation between $1/D_p$ and $d/\sigma$ still works very well for fixed crosslink density, although slight quantitative deviations are noticeable in **Figure 3a**. Moreover, we find that the exponent of the power law (i.e., the slope $k_{\text{size}}$ of the $\log(1/D_p)$ versus $d/\sigma$ plots) varies *linearly* with crosslink fraction $f_n$ (see **Figure S8**) to within slight uncertainties. This interesting result can be understood by combining our above findings that $\log(1/D_p) \propto T_g(f_{\text{cross}})$ and $C^2 \propto T_g(f_n)$ with the definition of confinement parameter ($C \sim f_{\text{cross}}^{1/2}$), as shown in the SI.

Overall, the SCCH theory calculations performed at the experimental measurement temperature agree well with the corresponding experimental and simulation results in **Figures 3a** and **3b**, respectively. Moreover, the predicted trends remain robust at elevated temperatures (see **Figure S10**). We note in passing that an alternative linear relationship between $\log(D_p)$ and $d/\sigma$ (an exponential, not power law, dependence) is found to be valid only over a more limited lower $d/\sigma$ range (see **Figure S9**), as expected [35, 64, 65].

### C. Confinement Mesh Perspective

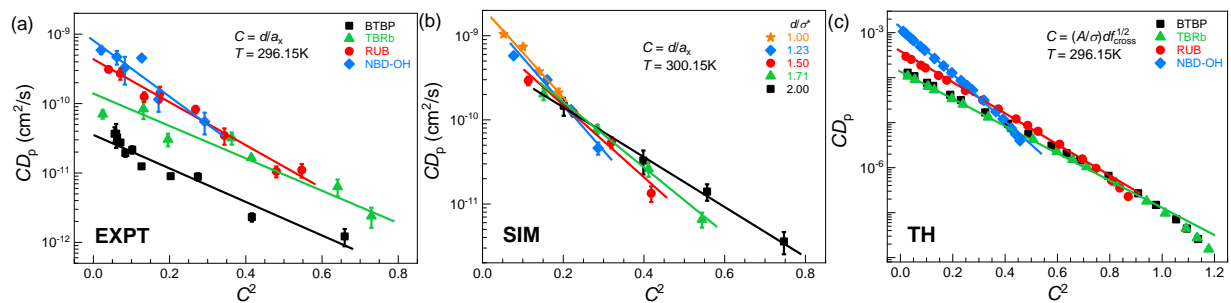

**Figure 4.** Penetrant diffusivity from an entropic mesh confinement perspective per eq(1). Data for various penetrants are plotted as $\log(CD_p)$ versus $C^2$ for: (a) EXPT, (b) SIM, and (c) TH over a wide range of crosslink fractions at a fixed temperature ($T = 296.15$K for EXPT and TH, and $T = 300.15$ K for SIM). The confinement parameter $C = d/a_x$ in EXPT and SIM, and $C =$

$(A/\sigma)d(f_n/0.19)^{1/2}$ in TH (see SI). The simulation $C$ is again multiplied by 0.573/1.03 to allow comparison to EXPT and TH.

Our analysis in the prior two sub-sections was motivated by the idea that the key factor in the determination of the penetrant diffusion constant is the crosslink-induced enhancement of its local activated hopping time. We now explore in detail the very different limiting perspective [per eqs(1) and (2)] that the entropic mesh confinement effect can empirically "explain" our results.

We first consider the theoretical results from an analytic perspective. Surprisingly, a "mesh confinement like" form can be obtained from an activated segmental dynamics analysis. This follows by combining the results $\log(\tau_{\alpha,p}/\tau_0) \propto T_g(f_n)$ and $C^2 \propto T_g(f_n)$ discussed above and in the SI, which implies the theory predicts $\log(\tau_{\alpha,p}/\tau_0) \propto C^2$. Hence, a *non-causal* "degeneracy of interpretation" at fixed temperature is predicted (not assumed) by the theory. Its origin is that crosslinking modifies the local penetrant hopping activation barrier in a manner that depends on the confinement parameter $C$ in the same manner as does the entropic mesh mechanism. In reality, an additional contribution to the penetrant barrier from a true entropic mesh confinement effect is potentially important, i.e., $X = \exp(-C^2)/C$ in $D_p = X\tau_0/\tau_{\alpha,p}$ of Eq (1), thereby yielding $CD_p \sim \exp(-BC^2)$, where $B$ is a $C$-independent numerical factor.

**Figure 4** aims to separate the consequences of the two very different physical effects on penetrant diffusivity and further test our new theoretical insights using experimental and simulation data. Linearity between $\log(CD_p)$ and $C^2$ is observed for all cases. However, the apparent slope in **Figure 4** decreases with penetrant size, which is the *opposite* trend found by varying penetrant size for the slopes in **Figures 2a-2c** which were constructed based on the very different crosslink fraction dependent activated segmental dynamics perspective. Our finding in **Figure S5** that $T_g(f_n) \propto C^2$ was deduced under fixed $d$ conditions, and thus the core insight is $T_\text{g}(f_\text{n}) \propto C^2/d^2$. Applying this proportionality to above analysis yields $\log(1/CD_\text{p}) \sim k_\text{cross} T_\text{g} \sim (k_\text{cross}/d^2)C^2$. **Figure 2d** shows that $k_\text{cross}$ varies very weakly (logarithmically) with penetrant size compared to the quadratic dependence in this expression, thereby providing an explanation for why the apparent slopes in **Figures 2a-2c** ($k_\text{cross}$) increase, but those in **Figure 4** ($B \sim k_\text{cross}/d^2$) decrease, with penetrant size. The theory results in **Figure 4c** also establish the operational robustness of the relation $CD_\text{p} \sim \exp(-BC^2)$ in the large $C$ regime, which is *not* due to entropic mesh confinement, but rather to crosslink dependent penetrant local activated hopping.

We have also tested whether the relation $D_\text{p} \sim \exp(-EC)$ with $E$ an adjustable constant [per Eq(2)] can empirically model our data. We find in the "weak confinement" range $0.15 < C < 0.9$ it works well for experiment, simulation, and theory (see **Figure S11**). This can be theoretically understood from an activated segmental relaxation perspective since only the local cage barrier is important in this regime and SCCH theory predicts its scales *linearly* with $C$. Hence, the success of the form $D_\text{p} \sim \exp(-EC)$ signals another non-casual "degeneracy of interpretation" of what controls penetrant diffusivity. More generally, by revisiting previous simulation data [42], we find that in a practical sense both relations motivated by an entropic mesh confinement dominated perspective (eqs(1) and (2)) work rather well in the limited range of $C$ analyzed (see **Figure S12**). However, in a large enough $C$ regime we predict based on SCCH theory such an empirical entropic mesh perspective fails (**Figure S11c**) due to the growing importance of longer-range collective elasticity of the polymer matrix in determining the penetrant hopping rate. Detailed discussion of these issues are presented in the SI, including the robustness of our new insights at elevated temperatures (see **Figure S13**).

### D. Mechanistic Theoretical Understanding

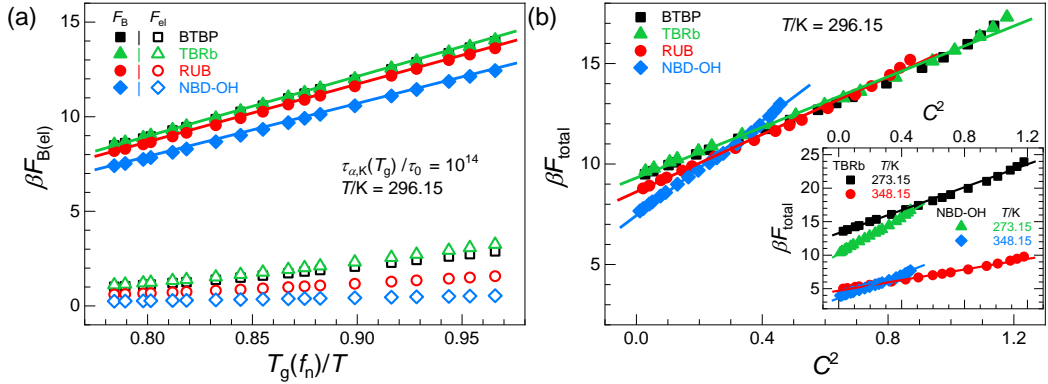

**Figure 5.** Theory predictions for the penetrant barriers in units of thermal energy. (a) Local cage and collective elastic barriers as a function of $T_g(f_n)/T$ over a wide range of crosslink fractions at a fixed $T = 296.15$K. The small differences between the elastic barriers of BTBP and TBRb reflect the fact their sizes are not exactly the same (see Table 1). (b) Main: Same as in (a) but for the total barriers as a function of $C^2$. Inset: Total barriers for TBRb and NBD-OH as a function of $C^2$ at a lower ($T = 273.15$K) and higher ($T = 348.15$K) temperatures.

The overall good agreement between experiment, simulation, and theory, despite being performed in non-identical supercooled regimes, suggests an intriguing simplicity is at play. In this sub-section, the physical mechanism is elucidated based on SCCH theory and our understanding of the penetrant local cage and longer-range collective elastic activation barriers as a function of temperature, size ratio, and crosslinking fraction.

**Figure 5a** shows that the collective elastic barriers for all molecular penetrants studied are far smaller than their local cage analogs. Regardless of the precise $T_g(f_n)$ vitrification criteria adopted, *both* the local cage and elastic barriers vary linearly with $T_g(f_n)/T$ over the entire $T_g/T$ range studied for all the penetrants. This provides a theoretical explanation of the exponential relationship between alpha relaxation time or diffusion constant and $T_g(f_n)/T$ in **Figure 2c**. Furthermore, although **Figure 5a** shows that the local cage and elastic barriers both increase with penetrant size, the rate of growth of the local cage barriers remain nearly independent of penetrant size while that for the elastic barrier strongly increases. Thus, the slope change in **Figure 2c** of log ($\tau_0/\tau_{\alpha,p}$) versus $T_g(f_n)/T$ plots with penetrant size is mainly a consequence of the increasing relevance of the longer-range collective elastic barrier relative to its local cage analog.

Combining the above results with $C^2 \propto T_g(f_n)$ discussed above and in the SI, we predict that the penetrant local cage and elastic barriers vary linearly with $C^2$ (see **Figure S14**). This provides an understanding of our numerical finding of a linear relationship between the penetrant total barrier and $C^2$ seen in **Figure 5b** and its inset, for various penetrant sizes and temperatures. This insight thus establishes the underlying physical reason for the operational usefulness of the relation $\tau_{\alpha,p}/\tau_0 \sim \exp[-(B-1)C^2]$ or $CD_p \sim \exp(-BC^2)$ (where $B$ is a numerical factor) even under deeply supercooled conditions where the penetrant total barrier is high. This prediction can be tested in future experiments and/or simulations that probe even slower mass transport characterized by higher total barriers than discussed in our present work.

Finally, because penetrant barriers are predicted to grow linearly with $C^2$, a literal linear relationship between them and $C$ does not apply, as confirmed in **Figure S14b**. However, when $C$ is not sufficiently large ($0.15 < C < 0.9$) corresponding to weak or non-existent mesh confinement, the penetrant local cage barrier does effectively grow linearly with $C$ when the collective elastic barrier is negligible. This explains why our experimental and simulation data (and also other literature data [42]) can be empirically fit by the exponential relation $D_p \sim \exp(-EC)$ over the limited range of $C$ measured. If in future experiments and simulations the confinement parameter is further increased and/or the temperature lowered to a more deeply supercooled regime, we predict the exponential relation $D_p \sim \exp(-EC)$ will eventually fail.

**Discussion and Conclusions**

We have combined experiment, simulation, and theory to unravel the competing effects of penetrant size on its transport in crosslinked polymer networks over a range of crosslink densities, size ratios, and temperatures. The crucial importance of the temperature and chemistry specific degree to which penetrant hopping is coupled to the polymer structural relaxation process has been established, with the entropic barrier due to entropic mesh confinement of quantitative, but secondary, importance. The leading order effect of network crosslinks is to slow down polymer structural relaxation and greatly suppress the elementary penetrant hopping event. However, there are also non-causal "empirical correlations" between the behavior of penetrant transport largely controlled by crosslink-induced slowing down of segmental relaxation and what is expected based on a mesh confinement perspective. The reason for this "degeneracy of interpretation" is understood within the theory as a consequence of a common dependence of the dynamic activation barriers on the confinement parameter $C$ and how the entropic obstruction barrier scales with $C$.

The good agreement between experiment, simulation, and theory provides strong support for the size ratio (penetrant diameter to Kuhn length) as a key variable, and more generally the usefulness of coarse-grained models that average over Angstrom scale chemical details in crosslinked networks. The SCCH theory reveals the microscopic physics underlying the behaviors found in experiment and simulation in terms of local cage and longer-range collective elastic barriers to penetrant hopping, and how they depend differently on the degree of crosslinking and temperature. Testable predictions in a more deeply supercooled regime not probed in our present experiments or simulations were made, where penetrant transport is predicted to become even more size sensitive. We believe our work illustrates the generality of the physical mechanisms identified which are relevant for selective membrane design, and establish the usefulness of simulation even at temperatures well above those often probed in experiment.

Of course, the relative importance of the segmental relaxation and mesh confinement effects on penetrant diffusion will depend on temperature (e.g., rubbery vs supercooled regime) and the precise value of the mesh confinement parameter. Further work is needed in this direction, and our analysis provides a roadmap to anticipate and understand how the two primary consequences of chemical crosslinking impact penetrant mass transport. Other major future directions are to apply our combined experiment-theory-simulation approach to address the dynamical impact of penetrant-polymer specific attractive interactions, generalize the simulation and theory approaches to explicitly treat the role of molecular penetrant shape and flexibility on transport, and address penetrant diffusion in associating polymer liquids [66] and vitrimers [67-70].

**Acknowledgements**. This research was supported by the U.S. Department of Energy, Office of Basic Energy Sciences, Division of Materials Sciences and Engineering, under Award No. DE-SC0020858, through the Materials Research Laboratory at the University of Illinois at Urbana−Champaign.

# Supplementary Information for "How Segmental Dynamics and Mesh Confinement Determine the Selective Diffusivity of Molecules in Dense Crosslinked Polymer Networks"

Baicheng Mei [a,d], Tsai-Wei Lin [c,d], Grant S. Sheridan [a,d], Christopher M. Evans [*a,c,d],

Charles E. Sing [*a,c,d] and Kenneth S. Schweizer [*a,b,c,d]

[a.] *Department of Materials Science, University of Illinois, Urbana, IL 61801, USA*

[b.] *Department of Chemistry, University of Illinois, Urbana, IL 61801, USA*

[c.] *Department of Chemical & Biomolecular Engineering, University of Illinois, Urbana, IL 61801, USA*

[d.] *Materials Research Laboratory, University of Illinois, Urbana, IL 61801, USA*

[*] *kschweiz@illinois.edu*    [*] *cesing@illinois.edu*    [*] *cme365@illinois.edu*

For the benefit of the reader, Section I recalls more details of the Materials and Methods underlying the experiments, simulations, and theory work in the main text. Section II provides additional theoretical and simulation evidence for the crucial role of network segmental relaxation in determining the penetrant diffusion constant at elevated temperatures, and results for different $T_g$ criteria to further establish the robustness of our findings. Section III presents additional simulation and theoretical results for the penetrant-segment size ratio dependence of the penetrant diffusion constant. The entropic mesh confinement analyses of the penetrant diffusion constant are further discussed in section IV. Finally, section V presents a more detailed discussion of the mechanistic insights deduced from SCCH theory.

## I. Details of Experiment, Simulation and Theory

### A. Experiment.

The synthesis of n-butyl acrylate (nBA) networks has been reported in our prior work.[1] In brief, the networks were synthesized via the neat free radical polymerization of nBA and 1,8-octanediacrylate with varying molar ratios from 0.01 to 0.5, using AIBN as the initiator at 65 °C with $10^{-5}$ M of four different dye molecules dissolved in the monomer. BTBP was the one dye studied previously [1], while the new three dyes of different size and shape are: tert-butyl substituted rubrene (TBRb, >99%, Ossila), rubrene (RUB, 99.99%, Sigma-Aldrich), and propanol substituted nitrobenzofurazan (NBD-OH, synthesized following a literature protocol [2]). The van der Waals volume of each penetrant was calculated using an atomic group contribution method [3,4]. These values are reported in **Table 1** along with the diameter corresponding to a sphere of equivalent volume. BTBP (aspect ratio ~3.5) and TBRb (more round plate-like shape) were selected in order to isolate the effect of penetrant shape since they have nearly identical volumes (**Table 1**). To consider the effect of different penetrant volumes, while fixing the aspect ratio, TBRb and RUB were used. The fourth penetrant NBD-OH is the most spherical molecule, has the smallest volume, and bears a positive and negative charge on the nitro group.

The emission/excitation spectra for each dye were determined using a Horiba PTI-Quantamaster to determine the concentration where aggregation was no longer a concern (**Figure S1**). Time-dependent illumination of the networks was used to evaluate the bleaching of the dye under similar conditions to the fluorescence (FRAP) experiments. The FRAP procedure for BTBP and analysis method were described in prior work [1]. For RUB, TBRb, and NBD-OH, the FRAP experiments were carried out using a 488nm laser on a Leica SP8 Confocal Microscope. The networks (~60 μm) were kept in an inert Argon glovebox on a glass coverslip until running FRAP. All measurements were performed at 23 °C on a Peltier temperature-controlled stage.

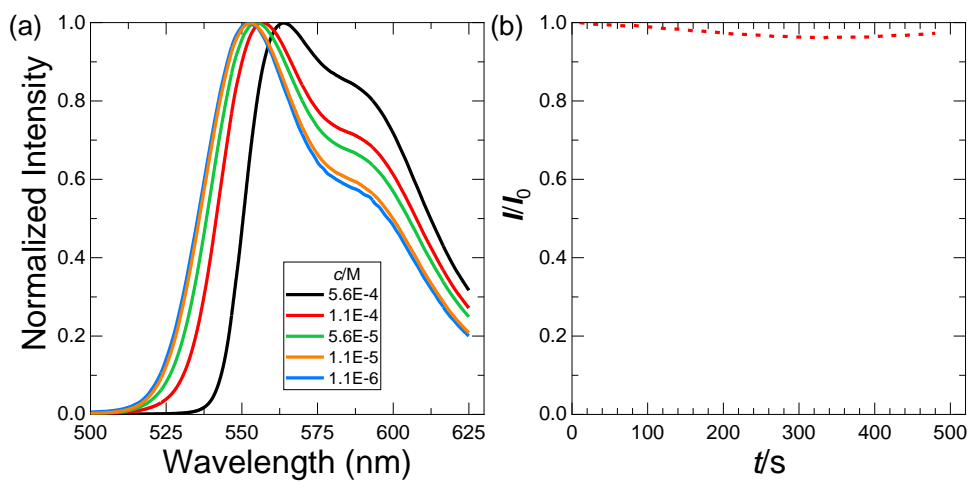

**Figure S1.** (a) Normalized RUB emission spectra intensity excited at 481 nm at various concentrations, $c$ (in unit of M), to determine the dilution needed for FRAP experiments. Below a $10^{-5}$ M dye concentration the emission spectra no longer changes (or changes very slightly). (b) RUB emission intensity stability at 481 nm excitation on the fluorimeter over 8 minutes showing only 5% bleaching indicating it is appropriate for the FRAP experiments.

The $T_g$ of each polymer network was determined by differential scanning calorimetry (TA Instruments DSC 2500) using a heat-cool-heat ramp protocol at 10 °C/min. The first derivative of the second heating curve was calculated to aid in visualizing the breadth which grows substantially with increasing crosslink density. Each of the dyes were incorporated into the DSC samples to confirm the dye does not impact the macroscopic physics of the networks at such trace levels. **Figure S2** shows the derivative heat flow curves for all networks in this study, and the BTBP samples were previously reported.[1] All networks show an increase in $T_g$ and the $T_g$ breadth with increased crosslinking, and these values are summarized in **Table S1**.

**Table S1**. Summary of DSC and DMA results for $T_g$, $T_g$ breadth $\Delta T_g$, $N_x$, and $a_x$ for the PnBA networks studied with the four types of dyes: (a) BTBP; (b) TBRb; (c) RUB; and (d) NBD-OH.

**(a) BTBP**

| Mol % crosslinker | $T_g$ (°C) | $\Delta T_g$ (°C) | $N_x$ | $a_x$ (nm) |
|---|---|---|---|---|
| 1.1 | -47 | 13 | 92 | 4.7 |
| 1.2 | -47 | 13 | 85 | 4.6 |
| 1.2 | -46 | 13 | 82 | 4.5 |
| 1.4 | -46 | 13 | 71 | 4.2 |
| 1.7 | -43 | 13 | 59 | 3.9 |
| 2.1 | -43 | 13 | 48 | 3.5 |
| 2.6 | -41 | 14 | 38 | 3.2 |
| 4.4 | -37 | 14 | 22 | 2.5 |
| 6.1 | -33 | 15 | 15 | 2.1 |
| 9.9 | -25 | 18 | 9.1 | 1.7 |
| 16.9 | -17 | 21 | 4.9 | 1.4 |

(b) TBRb

| Mol % crosslinker | $T_g$ (°C) | $\Delta T_g$ (°C) | $N_x$ | $a_x$ (nm) |
|---|---|---|---|---|
| 1.0 | -48 | 11 | 97 | 4.9 |
| 2.6 | -41 | 11 | 37 | 3.1 |
| 4.4 | -39 | 13 | 21 | 2.5 |
| 8.1 | -31 | 13 | 11 | 1.9 |
| 9.8 | -30 | 13 | 9.2 | 1.7 |
| 15.7 | -22 | 14 | 5.4 | 1.4 |
| 21.1 | -13 | 17 | 3.7 | 1.3 |

(c) RUB

| Mol % crosslinker | $T_g$ (°C) | $\Delta T_g$ (°C) | $N_x$ | $a_x$ (nm) |
|---|---|---|---|---|
| 1.0 | -47 | 11 | 99 | 4.9 |
| 1.9 | -45 | 11 | 53 | 3.7 |
| 3.7 | -40 | 11 | 26 | 2.7 |
| 5.0 | -38 | 12 | 19 | 2.3 |
| 8.1 | -32 | 13 | 11 | 1.9 |
| 10.8 | -28 | 14 | 8.3 | 1.7 |
| 16.0 | -19 | 17 | 5.3 | 1.4 |
| 18.6 | -18 | 14 | 4.4 | 1.3 |

(d) NBD-OH

| Mol % crosslinker | $T_g$ (°C) | $\Delta T_g$ (°C) | $N_x$ | $a_x$ (nm) |
|---|---|---|---|---|
| 1.0 | -47 | 10 | 100 | 4.9 |
| 3.3 | -41 | 11 | 30 | 2.8 |
| 4.5 | -36 | 11 | 21 | 2.5 |
| 7.3 | -33 | 12 | 13 | 2.0 |
| 10.3 | -28 | 13 | 8.7 | 1.7 |
| 19.3 | -11 | 16 | 4.2 | 1.3 |

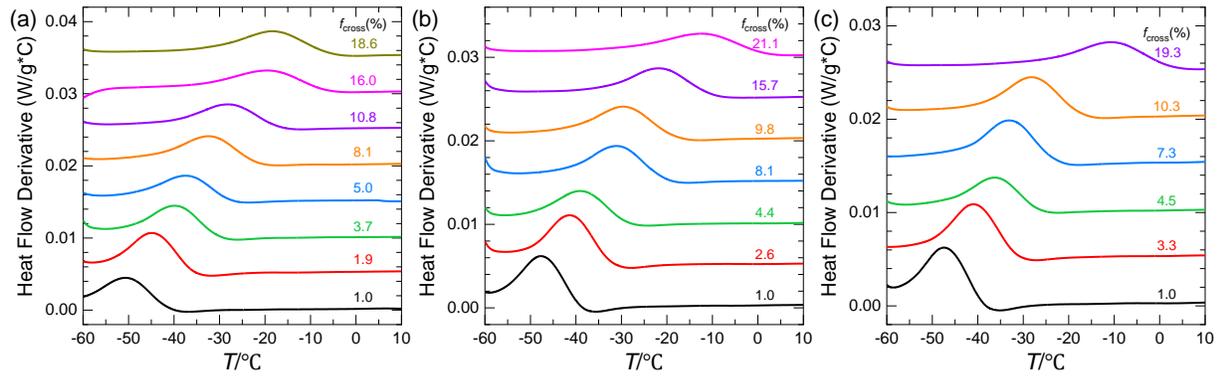

**Figure S2.** DSC derivative heat flow curves for PnBA networks with added dyes (a) RUB, (b) NBD-OH, and (c) TBRb. The $f_{cross}$ value is shown next to each curve. Dye incorporation does not affect the $T_g$ value or the dependence on crosslink density within errors.

Dynamic mechanical analysis (TA Instruments DMA Q800) was used to determine the elastic modulus of the networks which was then used to calculate the mesh size. Stress-strain curves are shown in **Figure S3**, and all exhibit a linear response up to 1.0 % strain. The slope is taken as the Young's modulus, and the presence of trace dye does not affect the value.

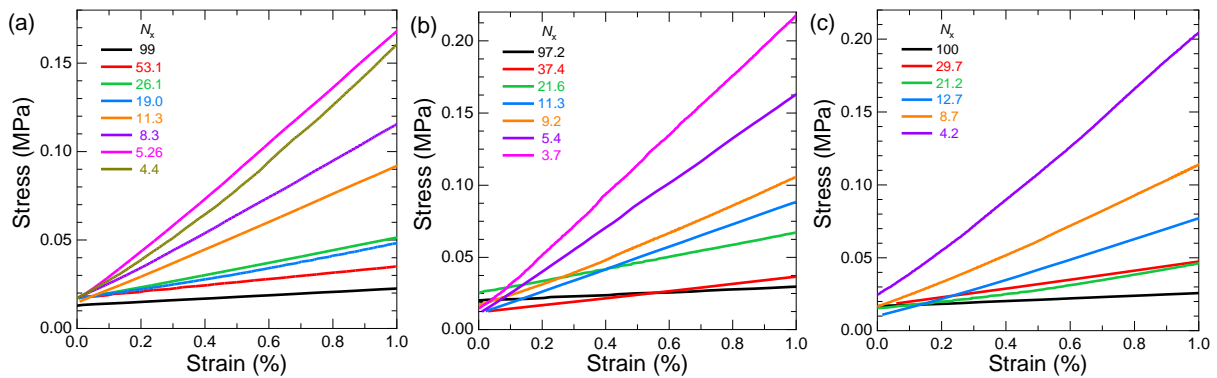

**Figure S3.** DMA stress-strain curves for networks with (a) RUB, (b) NBD-OH, and (c) TBRb taken at 23 °C under tension. Number average degree of polymerization between crosslinks is shown in the legends.

### B. Simulation

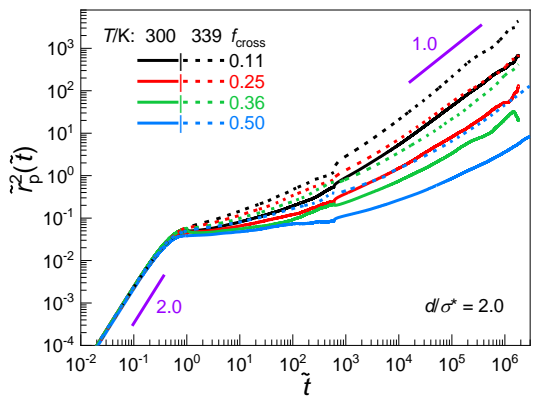

**Figure S4**: Time evolution of mean-square-displacement of penetrant, $\langle \tilde{r}_p^2(\tilde{t}) \rangle$, in simulation at different values of $f_{\text{cross}}$ at a lower (300K) and a higher (339K) temperatures. The penetrant diffusion coefficient is obtained in the diffusive regime (long-time scale) where the time scaling exponent is 1.0.

The simulation model is built upon our previous model for neat networks [5]. Here we briefly summarize our previous model and include simulation details for the penetrant diffusion in networks problem. Dimensionless simulation quantities with tildes (e.g. $\tilde{T} = T/T^*$) are employed to distinguish them from the analogous experimental and theoretical quantities. The following dimensionless variables are employed: bead diameter $\sigma^*$ for length (different from the $\sigma$ used in theory), temperature normalized by $T^*$ (the temperature scale determined from parametrization step, see details in ref. [5]), energies normalized by $k_B T^*$, and times normalized by $\tau^* = \sqrt{m\sigma^2/k_B T^*}$, where $m$ is the nBA monomer mass.

The initial system is composed of twenty linear chains ($N_c = 20$) with $N_m = 30$ the number of monomer beads per chain and $N_p$ the number of spherical penetrants of diameter $\tilde{d}$ in a cubic box with periodic boundary conditions in three dimensions. Polymers are modeled as standard semiflexible chains [6-11] where each bead represents a single nBA monomer.

The overall energy contains contributions from bonding interactions $\widetilde{U}_B$, a bending potential $\widetilde{U}_\theta$, and Lennard-Jones interactions $\widetilde{U}_{LJ}$:

$$\widetilde{U} = \widetilde{U}_B + \widetilde{U}_\theta + \widetilde{U}_{LJ} = \sum_{i>1} \tilde{u}_{B,i} + \sum_{i>1} \tilde{u}_{\theta,i} + \sum_{ij} \tilde{u}_{LJ,ij} \quad (S1)$$

Here the total system energy is written in terms of the pairwise contributions $\tilde{u}_{B,i}$, $\tilde{u}_{\theta,i}$, and $\tilde{u}_{LJ,ij}$ corresponding to their respective energy contributions. Bonded monomers interact through harmonic bonding potential:

$$\tilde{u}_{B,i} = \frac{\tilde{k}}{2}(\tilde{r}_{i,i-1} - 1)^2 \quad (S2)$$

where the large spring constant $\tilde{k} = 2000$ [12-15] is adopted to enforce the equilibrium distance between the $i$ and $i-1$ monomers as $\tilde{r}_{i,i-1} = 1$. A bending energy is introduced to account for chain stiffness:

$$\tilde{u}_{\theta,i} = \tilde{k}_\theta [1 - \cos\theta_i] \quad (S3)$$

where the bending constant $\tilde{k}_\theta = 1.52$ is selected to reflect the experimental Kuhn length of PnBA. Further details of the parametrization is given in ref [5]. A Lennard-Jones (LJ) potential is used to describe all non-bonded interactions,

$$\tilde{u}_{LJ,\alpha\beta} = \begin{cases} 4\tilde{\epsilon}_{\alpha\beta}\left[\left(\frac{\tilde{d}_\alpha+\tilde{d}_\beta}{2\tilde{r}_{\alpha\beta}}\right)^{12} - \left(\frac{\tilde{d}_\alpha+\tilde{d}_\beta}{2\tilde{r}_{\alpha\beta}}\right)^6\right], \tilde{r}_{\alpha\beta} < \tilde{r}_{cut} = 2.5 \times \frac{\tilde{d}_\alpha+\tilde{d}_\beta}{2} \\ 0, \text{ otherwise} \end{cases} \quad (S4)$$

where $\alpha, \beta \in \{n, p\}$, with n denoting the monomer bead in the network and p denotes the penetrant, and $\tilde{d}_n$ is always unity (i.e., $d_n = \sigma^*$) and $\tilde{d}_p \equiv \tilde{d}$. In this study, we only consider the effect of size of different penetrants (not specific penetrant-polymer attractions) so that $\tilde{\epsilon}_{nn} = \tilde{\epsilon}_{np} = \tilde{\epsilon}_{pp} = 1$.

The system is first equilibrated at $\tilde{T} = 1$, $\tilde{P} = 0$, and then networks are prepared by crosslinking the linear chains with reactive beads randomly distributed along the chain [16, 17]. The total number of reactive beads is $N_r = f_r N_m N_c$, where $f_r$ is the fraction of reactive beads which tunes the crosslink density of networks. If the distance between a reactive bead and a free bead (orange bead in **Figure 1b** of the main text) is within $\tilde{R}_{min} = 1.1$, a new permanent bond will be formed given an assigned probability. Once a reactive bead forms a bond with a free bead, both the reactive bead and the free bead are labeled as crosslink beads (dark blue beads in **Figure 1b** of the main text) and these two beads represent one crosslink 'molecule'. Crosslinking reactions were turned off once every reactive bead has formed a new bond with another free monomer (maximum number of possible bonds has been reached). In the end, $N_r$ reactive beads have reacted with the $N_r$ free beads and they turned into $2N_r$ crosslink beads and belong to the $N_r$ crosslink molecules.

The crosslink density, $f_{cross}$, of networks is defined as

$$f_{cross} = \frac{n_{crosslink}}{n_{crosslink} + n_{monomer}} = \frac{N_r}{N_m N_c - N_r}. \tag{S5}$$

Here, $n_{crosslink}$ and $n_{monomer}$ are the number of moles of crosslink molecules and nBA monomers. Four values of crosslink fraction $f_{cross}$ are considered: 0.11, 025, 0.36, and 0.5. The mesh size of networks at different crosslink densities is defined as the averaged distance between two adjacent crosslink beads on the same chain. The glass transition temperature $T_g$ is determined from a dynamic criterion. The alpha relaxation time ($\tilde{\tau}_\alpha$) at each temperature was defined as the time where the temporal autocorrelation function of a Kuhn monomer vector, $C_\lambda(\tilde{t}) = \langle P_2(\tilde{r}_3(\tilde{t}) \cdot \tilde{r}_3(0)) \rangle$ decays to $1/e$. [18-22] Here, $P_2$ is the second Legendre polynomial and $\tilde{r}_3$ is a vector between two beads that are 3 bonds apart which reflects the choice of coarse-grained bead relate to Kuhn segment of PnBA. $T_g$ is then defined as when the alpha relaxation time is $\tilde{\tau}_\alpha(\tilde{T}_g) = 10^5$. More information can be found in our previous work [5].

After crosslinking, the system is cooled to a target temperature at a cooling rate $\tilde{\Gamma} = 8.3 \times 10^{-6}$ (corresponding to $\Gamma = 1.25 \times 10^9$ K/s in experimental units) and further equilibrated at constant $\tilde{P} = 0$. Another short NPT run was performed and the mean volume $V$ is measured. We then switch to a NVT ensemble by setting the system volume to the mean volume $V$ and equilibrate the system before final production run at NVT. [23] All simulations are performed in LAMMPS [24] with a standard Nosé-Hoover thermostat and barostat [25-27].

### C. Theory

We here generalize for the first time the SCCH theory of dilute hard sphere (HS) penetrant diffusion in polymer melts [28] to crosslinked polymer networks (see **Figure 1c** in the main text for an illustration of the key conceptual elements). Within this theory [28-31], the activation barrier and corresponding mean hopping time for penetrant motion, and the extent of coupling of its transport with the early, medium, and late stages of the matrix structural relaxation process, can be predicted based on two *coupled* dynamic free energies for penetrant and matrix, respectively (see **Figure 1c** of the main text). In the dilute penetrant limit of interest here, the polymer matrix dynamics is described by its own dynamic free energy calculated based on bulk ECNLE theory [32-34] as recently extended to crosslinked polymer networks [5]. A tangent Koyama [35, 36] semiflexible chain model is adopted where polymer beads interact with beads on other polymers and with a spherical penetrant via a site-site hard-core repulsion (**Figure 1c** of the main text). The structural correlations that enter the dynamical theories are computed using the polymer reference interaction site model (PRISM) integral equation theory [5, 28, 34].

Per prior work [5], this athermal model is related to real thermal polymer networks by employing a well-known mapping procedure that equates the dimensionless compressibility computed from PRISM theory to its temperature dependent experimental analog obtained from PnBA equation-of-state data at 1 atm [5, 37]. This produces an a priori relation between packing fraction and absolute temperature. Crosslinking is modeled dynamically as immobile (pinned strictly or vibrate locally) bare beads of the Koyama semiflexible chain distributed in a regular manner along a chain per the schematic in **Figure 1c** of the main text. Pinning is taken to not modify the structural correlations relative to the baseline polymer melt, consistent with recent simulation findings [38]. Prior quantitative comparisons of the predictions of ECNLE theory for the alpha time of Kuhn segments (deduced from a Kramers mean first passage time calculation [39, 40] including collective elastic effects [5, 34]) in a PnBA crosslink network show excellent agreement with experiment and simulation [5]. This provides the foundational input required in SCCH theory [28-31] to construct the penetrant dynamic free energy which is coupled with the matrix dynamic free energy (see **Figure 1c** of the main text).

All technical details and equations concerning the implementation of SCCH theory can be found in the literature [28-31]. Briefly, the penetrant alpha process is of a coupled local-nonlocal spatial character, involving hopping over a local cage barrier ($F_{B,p}$) due to its local interactions due to neighboring polymer matrix particles (Kuhn segments), which is coupled with small, but long range, collective displacements of all matrix particles outside the cage characterized by an elastic barrier ($F_{el,p}$). When the temperature is relatively high (weakly supercooled regime) or the penetrant-to-matrix size ratio is small, the elastic barrier is negligible, and the local cage barrier dominates the penetrant hopping or alpha relaxation process. But in the deeply supercooled regime for sufficiently large penetrants, both the local cage and elastic barrier are crucial. The theory self-consistently predicts the magnitude of facilitating activated dynamic fluctuations of the matrix that are required to achieve a penetrant hop over its barrier, yielding a dependence of the penetrant activated relaxation time on the crosslink density, penetrant-to-matrix size ratio, and temperature. SCCH theory focuses on the local penetrant hopping event and does not explicitly address geometric mesh confinement effects on penetrant diffusivity.

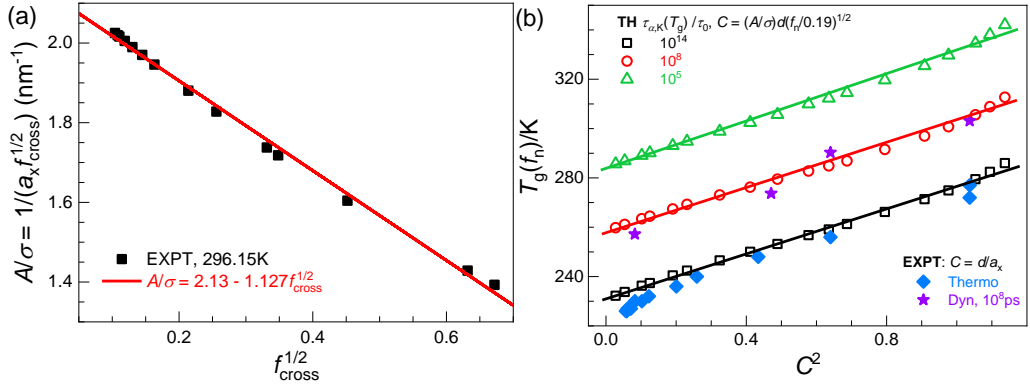

**Figure S5**. (a) The prefactor $A/\sigma = 1/(a_x f_{\text{cross}}^{1/2})$ in units of inverse nm for PnBA networks as a function of the square root of crosslink density $f_{\text{cross}} \equiv f_n/0.19$. The red line is a linear fit, $A/\sigma = 2.13 - 1.127(f_n/0.19)^{1/2}$, that is used as input in TH to calculate the confinement parameter as $C = (A/\sigma)d(f_n/0.19)^{1/2}$ and the corresponding penetrant diffusion constant. Because $f_{\text{cross}}$ and $a_x$ are negligibly affected by temperature, the fitting formula has no temperature dependence. (b) Glass transition temperatures defined based on different timescale criteria in TH and EXPT as a function of squared confinement parameter, $C^2$. Note, $C$ is computed based on $C = (A/\sigma)d(f_n/0.19)^{1/2}$ and $C = d/a_x$ in the TH and EXPT, respectively.

In experiments and simulations, one can define the confinement parameter as $C = d/a_x$ and compute the mesh size directly, as discussed previously [23, 41]. However, this approach is not directly applicable in the theory since our modeling of crosslinks via the neutral pinning of beads along the polymer chain does not directly allow the evaluation of $a_x$. In experiments, the mesh size $a_x$ is assumed proportional to the square root of the mean number of bare beads between two neighboring crosslinks $N_x$, i.e., $a_x \propto \sigma N_x^{1/2}$. Based on our recent study of PnBA crosslinked networks [5], the experimental crosslink fraction was defined as $f_{\text{cross}} = 1/N_x$ which is used in the present work. As such, one obtains $a_x \propto \sigma/f_{\text{cross}}^{1/2}$, and hence we take the confinement parameter to be $C = d/a_x = A(d/\sigma)f_{\text{cross}}^{1/2}$, where the prefactor $A$ is the proportionality factor between $a_x$ and $\sigma/f_{\text{cross}}^{1/2}$. Thus, we obtain $A/\sigma = 1/(a_x f_{\text{cross}}^{1/2})$. Based on purely experimental results for $a_x$ and $f_{\text{cross}}$, we compute $A/\sigma$ and plot it as a function of the square root of crosslink fraction $f_{\text{cross}}^{1/2}$ in **Figure S5a**. Previous simulations for semi-dilute crosslinked polymer solutions [23] also adopted $N_x$ to evaluate the mesh size and introduced a constant prefactor of 1.94 in defining the confinement parameter, which is the analog of the $A$ parameter. Using the estimate of a bare bead size for PnBA as 1 nm, one obtains $A/\sigma = 1.94$ nm$^{-1}$. In **Figure S5a**, we find the change of $A/\sigma$ is very limited, decreasing from 2.0 to 1.4 in unit of nm$^{-1}$, consistent in magnitude with 1.94 nm$^{-1}$ but our $A/\sigma$ does have a weak crosslink fraction dependence. An apparent linear behavior is observed between $A/\sigma$ and $f_{\text{cross}}^{1/2}$ (see red line in **Figure S5a**), given as $A/\sigma$ (nm$^{-1}$) = $2.13 - 1.127 f_{\text{cross}}^{1/2}$. Using this $A/\sigma$ (nm$^{-1}$) as input, we define the confinement parameter in our theoretical analysis as $C = [2.13 - 1.127(f_{\text{n}}/0.19)^{1/2}]d(f_{\text{n}}/0.19)^{1/2}$, where the penetrant diameter $d$ for real systems is given in **Table 1** of the main text. Importantly, we know that $f_{\text{cross}} = f_{\text{n}}/0.19$ from our previous quantitative study of pure PnBA melts and crosslinked networks [5], and empirically find that the crosslink density $f_{\text{n}}$ in our theory is proportional to $f_{\text{cross}}$ in experiment with a multiplier of 0.19.

## II. Role of Segmental Relaxation: $D_p$ versus $T_g(f_n)$

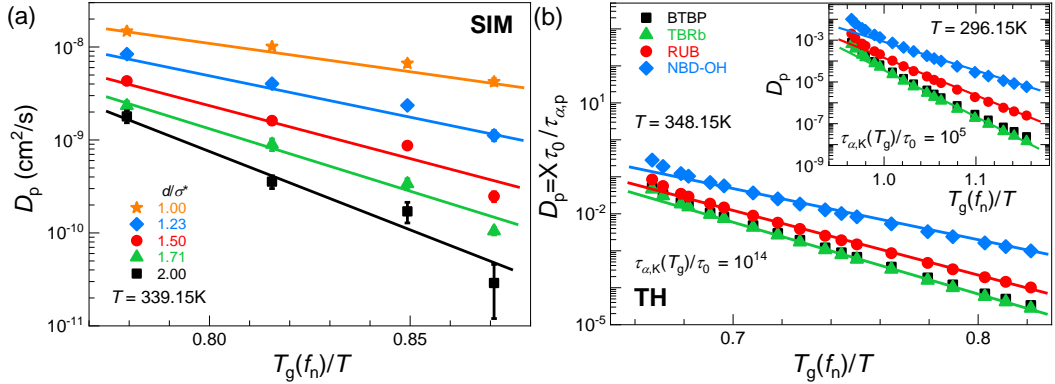

**Figure S6.** Relationship between penetrant diffusivity and $T_g(f_n)$ at higher temperatures than studied in EXPT. Diffusion constants for the different size penetrants are plotted as a function of $T_g(f_n)/T$ over a wide range of crosslink fractions for temperatures different than **Figure 2** in the main text for (a) SIM at $T = 339.15$K and (b) TH at $T = 348.15$K, with $D_p = X(\tau_0/\tau_{\alpha,p})$ in the TH and $X \equiv \frac{\exp(-C^2)}{C}$. In the inset of (b), we also show $D_p$ at $T = 296.15$K as a function of $T_g(f_n)/T$ based on different $T_g$ criteria: $\tau_{\alpha,K}(T_g)/\tau_0 = 10^5$ (essentially identical to that used in the simulations) which is different from that in the main frame where $\tau_{\alpha,K}(T_g)/\tau_0 = 10^{14}$ is adopted

in the spirit of the typical experimental vitrification definition that the polymer alpha relaxation time equals 100 s.

Taking the diffusion constant as a product of the inverse alpha time $\tau_0/\tau_{\alpha,p}$ and the direct entropic mesh confinement contribution formula $X = \exp(-C^2)/C$, i.e., Eq (1) of the main text with $b = 1$, we can theoretically calculate the probe diffusivity for the four organic dyes as shown in **Figure 2c** of the main text (solid symbols). One finds the combination of activated segmental relaxation physics (i.e., the penetrant hopping time contribution originates from the coupling of its motion with the segmental relaxation process) and entropic mesh confinement effects does not modify the exponential form of the relationship (penetrant diffusion constant or inverse alpha time versus $T_g/T$), although the corresponding apparent activation energy quantitatively increases (see **Figure 2d** of the main text) upon including the mesh confinement contribution as it must. When varying the temperature, the entropic confinement parameter contribution remains unchanged, while the activated hopping contribution changes a lot. This is shown in **Figure S6b** for the theoretical predictions at a higher temperature where one sees that the exponential form of the relationship remains unchanged, but the slope decreases with temperature, consistent with the simulation findings in **Figure 2b** of the main text and **Figure S6a**.

Finally, **Figure S7** presents a plot of the effective activation barrier, $k_{cross}$, versus the penetrant-to-matrix size ratio, $d/\sigma$, from **Figure 2d** of the main text in a log-log manner. One sees it also exhibits rather good linearity. This suggests an alternative power law behavior that differs from the logarithmic relationship in **Figure 2d** is also reasonable.

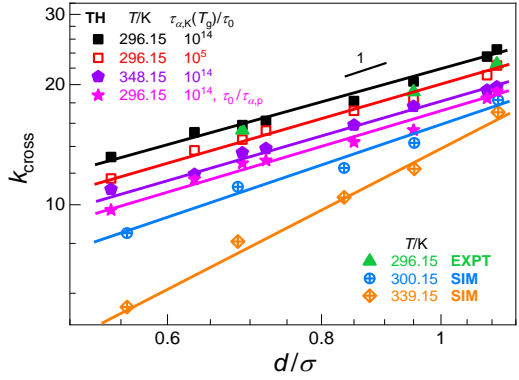

**Figure S7.** Effective activation energies per an "apparent" Arrhenius behavior, i.e., the slope of $\log(1/D_p)$ versus $T_g(f_n)/T$, as a function of size ratio $d/\sigma$ plotted in a *log-log* manner for different temperatures and methods (EXPT, SIM, TH). For the theoretical calculation, we also present the corresponding slope deduced from a $\log(\tau_{\alpha,p}/\tau_0)$ versus $T_g(f_n)/T$ plot. Note: as in **Figure 2d** of the main text, the x-axes of SIM data is multiplied by the factor 0.573/1.03.

### III. Penetrant-to-Matrix Size Ratio Dependence

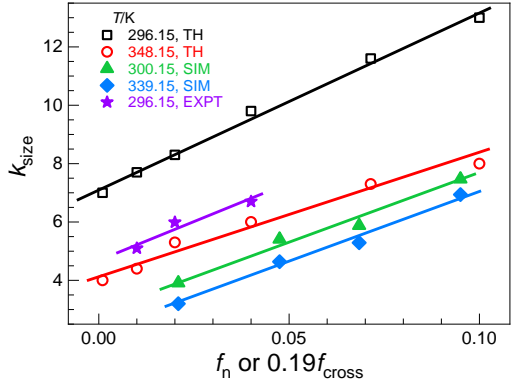

**Figure S8.** Linear-linear plot of the crosslink density dependence of the power law exponent for the $D_p$ versus $d/\sigma$ plot, i.e., the slope of $\log(1/D_p)$ versus $\log(d/\sigma)$, in EXPT, TH and SIM at various temperatures.

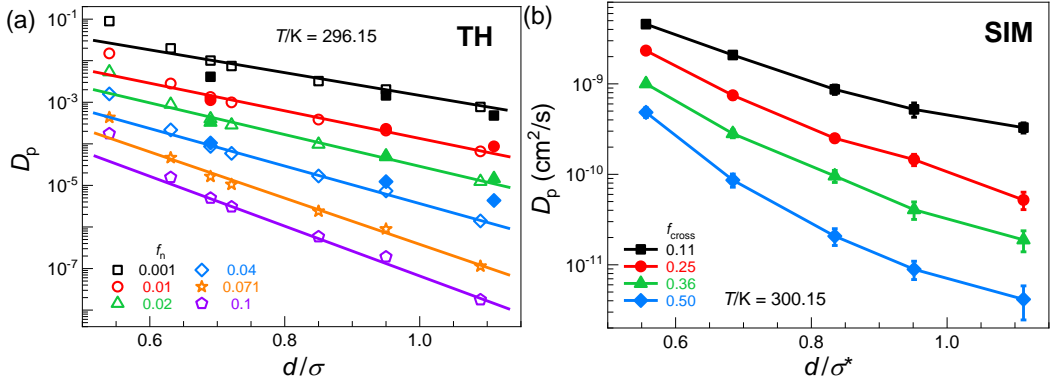

**Figure S9.** Penetrant size ratio dependence of its diffusivity plotted in log-linear manner at various fixed crosslink densities as a function of size ratio $d/\sigma$ in (a) TH at $T = 296.15\,\text{K}$ and (b) SIM at $T = 300.15$ K. The simulation value of $d/\sigma$ is multiplied by the factor of $0.573/1.03$ for comparison with the TH and EXPT results.

For the model of nanoparticle diffusion in rubbery polymer networks [23, 41, 42], when the confinement parameter is small enough there is no entropic barrier, and a power law relation between spherical particle diffusion constant and its diameter is predicted based on a hydrodynamic argument (Stokes relation). There are three differences between our SCCH theory and this approach: (i) Our prediction (and also our simulation and experimental results) applies solely to larger size penetrants; (ii) The physical relaxation mechanism underlying our power law relationship originates from penetrant activated hopping dynamics, not hydrodynamics in a non-activated regime where the confinement parameter is very low; (iii) The power law exponent for the model of ref. [41] is 3, generally much smaller than our crosslink density dependent exponent due to the qualitatively different physics (see **Figure S8**).

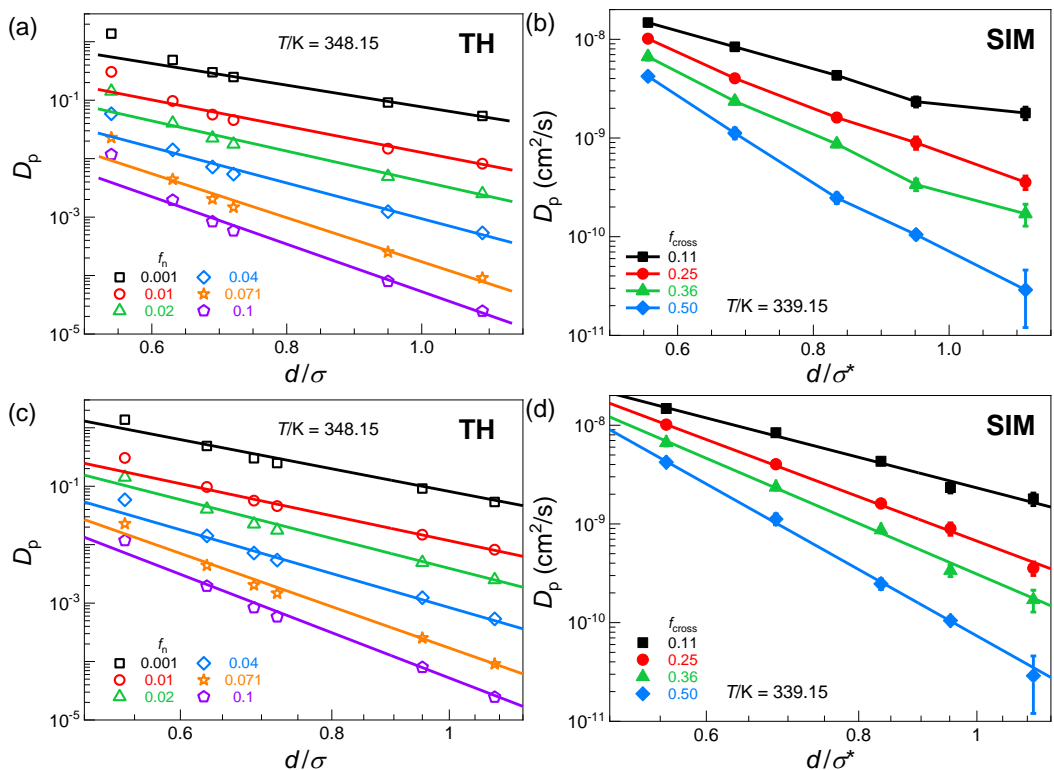

**Figure S10.** Penetrant size ratio dependence of its diffusivity at a higher temperature than studied experimentally. Same displays as **Figure S9** for (a) TH at $T = 348.15$ K and (b) SIM at $T =$

339.15K. The simulation value of $d/\sigma$ is again multiplied by the factor 0.573/1.03 for comparison with TH and EXPT. (c) and (d) are the same displays as (a) and (b), respectively, but in a log-log manner.

Although the exponential relationship between diffusion constant and penetrant-to-matrix size ratio in the smaller $d/\sigma < 0.5$ regime has been previously discussed and experimentally confirmed for chemical diverse penetrants in various polymer melts [37] (and in simulation [43]), to date the power law relationship lacks direct experimental and/or simulation evidence under variable crosslink density conditions. In the main text we provided a first simulation test of the power law relationship (since only the large $d/\sigma > 0.5$ regime is probed in our present work) between penetrant diffusivity and its size in crosslinked networks. The simulations find that for each crosslink density at a medium supercooled temperature ($T = 300.15K$) the predicted power law relationship is well obeyed (see **Figure 3b**). Moreover, the absolute value of the slope in the log-log plot (the power law exponent) increases linearly with crosslink density as shown in **Figure S8** where $\log(1/D_p) \propto T_g(f_{cross})$, $C^2 \propto T_g(f_n)$ and $C = A(d/\sigma)f_{cross}^{1/2}$ are combined and the very weak dependence of the prefactor $A$ on crosslink density is ignored (see **Figure S5a**, $A/\sigma$ only varies 30%, from 2.0 to 1.4 in unit of $nm^{-1}$, while $f_{cross}^{1/2}$ changes by a factor of ~7). This good agreement between simulation and theory provides additional support for the physical picture predicted by SCCH theory.

The question of an alternative linear relationship between $\log(D_p)$ and $d/\sigma$ is addressed in **Figure S9** for both the theory and simulation. It is found to be valid only over a limited range of low $d/\sigma$ ratios. Finally, the effect of temperature is considered in **Figure S10**, and our conclusions drawn at a single experimental temperature are found to remain robust.

## IV. Confinement Parameter Dependence

**Figure 4c** of the main text presented SCCH theory predictions in the plotting format $CD_p \sim \exp(-BC^2)$, which was shown to work rather well not only in the low $C$ regime, but also in the higher $C$ regime. This is nontrivial since in the low $C$ regime only local caging of a penetrant by the surrounding polymer segments is present, while in the larger $C$ regime both local caging and the facilitation effect of longer-range elastic cooperative motion of Kuhn segments is important. However, in *both* cases SCCH theory predicts the same $C-$dependence of the penetrant diffusion constant. This seems consistent with our experimental and simulation results in **Figures 4a** and **4b**, respectively, that find an Arrhenius-like dependence of the penetrant diffusion constant which we argue is because only the local cage barrier is significant (elastic barrier relatively small or negligible) for the present systems of interest, as seen in **Figure 5** and **Figure S14**.

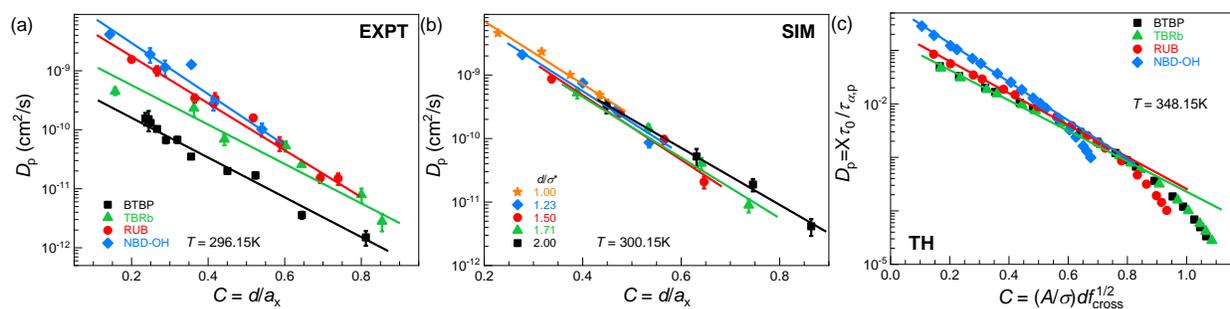

**Figure S11.** Log-linear plot of the penetrant diffusivity of various penetrants as a function of $C$ in (a) EXPT, (b) SIM, and (c) TH over a wide range of crosslink fractions at a fixed medium temperature ($T = 296.15$K for EXPT and TH, $T = 300.15$K for SIM). Here, $C = d/a_x$ in EXPT and SIM, and $C = (A/\sigma)d(f_n/0.19)^{1/2}$ in TH. The simulation confinement parameter $C$ is multiplied by the factor 0.573/1.0 for comparison with EXPT and TH.

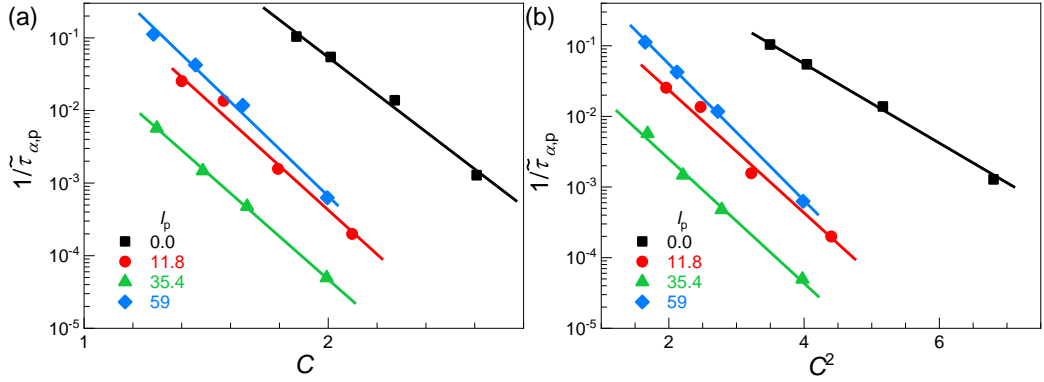

**Figure S12.** Literature simulation data [42] for the inverse penetrant alpha time as a function of the confinement parameter, plotted versus (a) $C$ and (b) $C^2$ for a crosslinked network of semiflexible polymers of very different persistent lengths $l_\mathrm{p}$ over a range of crosslink densities at a fixed penetrant size ($d = 6.4 r_\mathrm{c}$ with $r_\mathrm{c}$ being the simulation length unit).

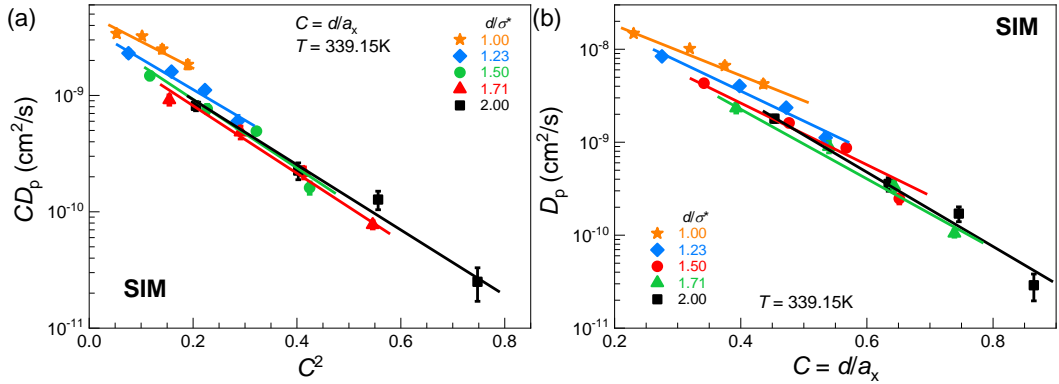

**Figure S13.** Relationship between penetrant diffusivity and confinement parameters at higher temperature. (a) and (b) SIM data in the same display format as **Figure 4** of the main text and **Figure S11b** adopted to test the relationships $D_\mathrm{p} \sim \exp(-EC)$ and $CD_\mathrm{p} \sim \exp(-BC^2)$, respectively, at a higher temperature $T = 339.15\mathrm{K}$.

**Figures S11a** and **S11b** present the corresponding results of $D_\mathrm{p}$ for experiment and simulation, respectively, for various penetrants as a function of the confinement parameter $C$ at an intermediate temperature. An exponential relationship works reasonably well in the limited data range of $0.15 < C < 0.9$, and essentially just as well as the relationship $CD_\mathrm{p} \sim \exp(-BC^2)$, signaling another "degeneracy of interpretation". The corresponding SCCH theory result is shown in **Figure S11c**, and in the same limited $C$ range an exponential relationship between of $D_\mathrm{p}$ and $C$ is predicted by the theory. However, the exponential relation breaks down (predicted curve bends down) beyond a sufficiently large confinement parameter. To understand this behavior, we plot the penetrant local cage and elastic barriers as a function of confinement parameter $C$ in **Figure S14b**. One sees that when $C$ is not too large the elastic barrier is negligible and penetrant dynamics is determined by local cage barrier which scales linearly with $C$. Thus, the $D_p \sim \exp(-EC)$ behavior is predicted to apply in the weakly or intermediately supercooled regime and/or for small size penetrants with $E$ a constant prefactor.

Finally, we find that all the above observations and predictions remain robust when changing temperature, as shown in **Figures S13a** and **S13b** for the simulation data which tests the relationships $CD_p \sim \exp(-BC^2)$ and $D_p \sim \exp(-EC)$, respectively, at a higher temperature. However, at a higher temperature, the activated hopping contribution to penetrant diffusivity does make a quantitatively smaller contribution relative to the entropic mesh confinement effects, as physically expected. This results in the slope $B$ discussed in the main text changing relatively weakly if the penetrant size is varied, as observed in our experiments, simulations, and theory calculations.

## V. Theoretical Barriers and Physical Mechanisms

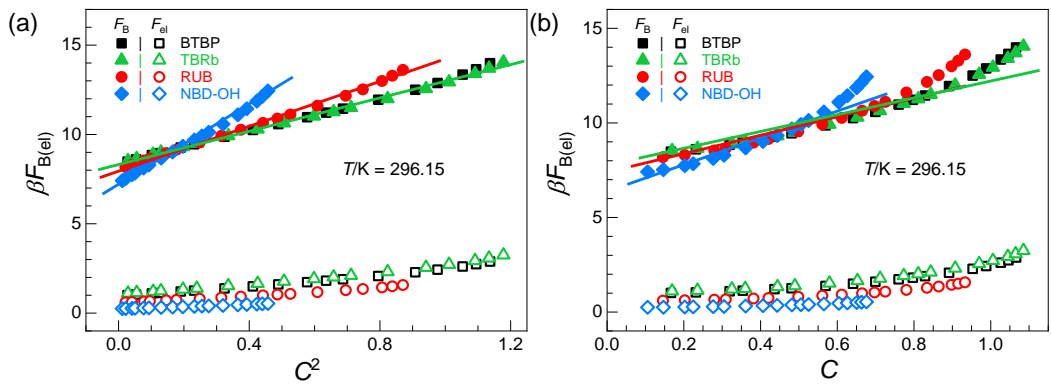

**Figure S14.** Theoretical predictions for the activation barriers in units of thermal energy. (a) Local cage and elastic barriers for the four molecular penetrants plotted as a function of $C^2$ with $C = (A/\sigma)d(f_n/0.19)^{1/2}$ over a wide range of crosslink fractions at a fixed medium temperature $T = 296.15$K. (b) Same display as in (a) but with $C^2$ replaced by confinement parameter $C$.

For all four molecular penetrants studied experimentally, their respective elastic barriers are predicted to be far smaller than their respective local cage barriers. There are several reasons for this: (i) compared to hard sphere fluids, for polymer systems the elastic barrier is relatively smaller because of the influence of chain connectivity as previously discussed [44]; (ii) the penetrant sizes (listed in **Table 1**) are relatively small compared to the Kuhn segment length of 1.72 nm, which is predicted in polymer melts [28] to result in rather strong decoupling between the penetrant and matrix dynamics and thus a low value of the elastic barrier; and (iii) crosslinking decreases the relative importance of elastic barrier compared to the local cage barrier. These effects collectively underlie the trends in **Figure 5a** of the main text and **Figure S14** where the elastic barrier increases much *slower* with $T_g(f_n)/T$, $C^2$ or $C$ relative to that of local cage barrier.

**References.**